\documentstyle[12pt]{article}
\newlength{\dinwidth}
\newlength{\dinmargin}
\begin{document}
\titlepage
\begin{flushright}
DTP/97/56 \\
July 1997 \\
\end{flushright}

\begin{center}
\vspace*{2cm}
{\Large \bf Single particle spectra in deep inelastic scattering
as a probe of small $x$ dynamics} \\
\vspace*{1cm}
J.\ Kwiecinski\footnote{On leave from H.\ Niewodniczanski
Institute of Nuclear Physics, Department of Theoretical Physics,
ul.\ Radzikowskiego 152, 31-342 Krakow, Poland.}, S.C.\ Lang,
A.D.\ Martin

\vspace*{0.5cm}

Department of Physics, University of Durham, Durham, DH1 3LE, UK.
\end{center}
\vspace*{1.5cm}

\vspace*{2cm}

\begin{abstract}
We study the transverse momentum ($p_T$) spectrum of charged
particles produced
in deep inelastic scattering (DIS) at small Bjorken $x$ in the
central region
between the current jet and the proton remnants.  We calculate
the spectrum at large $p_T$ with the BFKL $\ln (1/x)$ resummation
included and then repeat the calculation with it omitted.  We
find that data favour the former.  We normalize our BFKL
predictions
by comparing with HERA data for DIS containing a forward jet. 
The shape of the
$x$ distribution of DIS + jet data are also well described by
BFKL dynamics.
\end{abstract}

\newpage

\noindent {\large \bf 1.  Introduction}
\vspace*{0.3cm}\\
An intriguing feature of the measurements at HERA in the small
$x$ domain is the 
possible existence of significant $\ln (1/x)$ effects.  A major
part of the
rise observed
for the structure function $F_2$ with decreasing $x$ may be
attributed to the
resummation of the leading $\ln (1/x)$ \lq BFKL' \cite{BFKL}
contributions.
An excellent unified BFKL/GLAP fit of $F_2$ in the HERA regime
has recently been obtained using a \lq \lq flat in $x$" input
\cite{STASTO}, and the
rise due to BFKL-type effects has been quantified within this
description. However, the growth of 
$F_2$ with decreasing $x$ can be described equally well by pure 
GLAP \cite{DGLAP} $\ln (Q^2)$ 
evolution from suitably chosen input parton distributions so the
main origin of the rise is still an open question. The observable
$F_2$ 
is too inclusive to distinguish between these alternatives.  The
study of deep inelastic 
scattering (DIS) events containing an isolated forward jet 
\cite{M,KMS1} is a better discriminator of the underlying small
$x$ dynamics.  The process is sketched in
Fig.~1(a). In this case we effectively study DIS off known parton
distributions and so we 
avoid the ambiguity in the choice of the input distributions. 
The method is theoretically attractive.  The summation of the
leading $\ln
(1/x)$ contributions gives an $(x/x_j)^{-\lambda}$ behaviour of
the 
BFKL ladder connecting the photon to 
parton $a$.  Here $x$ is Bjorken $x$ and $x_j$ is the fraction of
the proton's 
longitudinal momentum carried by the parton jet. 
An unambiguous measurement of the exponent $\lambda$
looks feasible.  In practice a major problem is the
identification of the jet
due to parton $a$, and the measurement of its momentum, when it
is close to the remnants of the proton.  Typically the clean 
observation of the jet requires
$x_j \lower .7ex\hbox{$\;\stackrel{\textstyle <}{\sim}\;$} 0.1$ and so in 
this process we lose about a factor of 10
in the \lq small $x$ reach' of HERA.

Besides the $x^{-\lambda}$ growth as $x$ decreases along the BFKL
ladder, a 
second characteristic feature is the diffusion in $\ln k_T^2$
where $k_T$ are the transverse momenta of the gluons emitted
along the 
chain.  One way the diffusion
manifests itself is in an enhancement of the transverse energy
$(E_T)$ flow in the central region between the current jet and
the 
proton remnants \cite{gkms}, see Fig.~1(b).
In principle the diffusion can enhance $E_T$ from both the \lq
upper' and \lq lower' BFKL gluon ladders, which are denoted by
$\Phi$ and
$f$ in Fig.~1(b). However, the $x$ reach at HERA is insufficient
to 
fully develop the $\ln k_T^2$
diffusion in both ladders simultaneously.  Nevertheless, the
effect is quite
appreciable giving at the {\it parton} level an energy flow $E_T
\lower .7ex\hbox{$\;\stackrel{\textstyle <}{\sim}\;$} 2$ GeV/unit of rapidity. 
However the clean parton level
prediction can in practice 
be masked or mimicked by the effects of hadronization.  Thus,
although the prediction for $E_T$ is in agreement with
observations
\cite{ETflow} we
cannot definitely conclude that it is due to $\ln (1/x)$
resummations.

An interesting way to overcome this ambiguity is to consider the
emission of single particles at relatively large transverse
momentum $p_T$ 
in the central region \cite{MK}.
The single particle spectrum at sufficiently large values of
$p_T$ should be 
much more immune from hadronization and more directly reflect the
$\ln k_T^2$
diffusion from the BFKL ladders. 

The outline of the contents of the paper is as follows.  In
Sec.~2 we use the data for the process DIS + forward jet to
normalise the BFKL function $\Phi$  shown in Fig.~1(a).  To be
precise we numerically solve the BFKL equation for $\Phi$ using
the amplitude $\Phi^{(0)}$ for the quark box (and crossed box) as
input at a value $z_0$ of $z = x/x_j$ which is chosen so that the
resulting $\Phi$ reproduces the DIS + jet data.  Also, for
completeness, we present in Sec.~2, an analytic form for
$\Phi$ which is valid for fixed $\alpha_S$, and which has been
the basis for a recent analysis.  In Sec.~3 we give the formula
necessary to calculate the transverse momentum $(p_T)$ spectrum
of single particles produced in the central region.  The process
is shown in Fig.~1(c).  The predictions for the $p_T$ spectra
(with and without the BFKL effects included) are compared with
HERA data.  Finally in Sec.~4 we give our conclusions. \\

\noindent {\large \bf 2. DIS + forward jet events}
\vspace*{0.3cm}\\
We first calculate the cross section for DIS containing a forward
identified jet. 
This so-called \lq \lq Mueller" process is a valuable probe of
small $x$ dynamics
in its own right.  We compare with HERA data to normalise the
function $\Phi$
shown in Fig.~1(a).  There are uncertainties in the
normalisation, and even
the shape of the $x$ distribution is dependent on subleading $\ln
(1/x)$
corrections.

The variables of the process are shown in Fig.~2.  As usual the
variables $x$ and $y$ are given by $x = Q^2/2p \cdot q$ and $y =
p \cdot q / p_e
\cdot p$ where $p$, $p_e$ and $q$ denote the four momenta of the
proton, the incident electron and the virtual photon
respectively, and $Q^2 \equiv -q^2$.  The variables $x_j$ and
$k_{jT}$ are the
longitudinal momentum fraction and transverse momentum carried by
the forward jet.  The differential cross 
section is given by
\cite{M}
\begin{equation}
\frac{\partial \sigma_{j}}{\partial x \partial Q^{2}} = \int
dx_{j}
\int dk_{jT}^{2} \frac{4 \pi \alpha ^{2}}{xQ^{4}} \left[
\left(1-y \right) \frac{\partial F_{2}}{\partial x_{j} \partial
k_{jT}^{2}} + \frac{1}{2} y^{2} \frac{\partial F_{T}}{\partial
x_{j} \partial k_{jT}^{2}} \right] 
\label{eq:a2}
\end{equation}
where the differential structure functions have the following
 form
\begin{equation}
\frac{\partial^2 F_{i}}{\partial x_{j} \partial k_{jT}^{2}} =
\frac{3 \alpha_S (k_{jT}^{2}) }{\pi k_{jT}^2}
\sum_{a} f_a \left(x_{j},k_{jT}^{2} \right) \Phi_i
\left(\frac{x}{x_{j}},k_{jT}^{2},Q^{2} \right)
\label{eq:a3}
\end{equation}
for $i=T,L$ and $F_2=F_T+F_L$.  
We have assumed strong ordering at the parton $a$ -
gluon vertex. Assuming also $t$-channel pole dominance the sum
over 
the parton distributions is given by
\begin{equation}
\sum_{a} f_{a} = g+\frac{4}{9} \sum_q \left(q+ \bar q \right).
\label{eq:a4}
\end{equation}
Recall that these parton distributions
are to be evaluated at $(x_{j},k_{jT}^2)$ where 
they are well-known from the global analyses, so there are no
ambiguities arising from a non-perturbative input.\\

The functions $\Phi_i (x/x_j, k_{jT}^{2}, Q^2)$ 
describe the virtual $\gamma$ + virtual gluon fusion process 
including the ladder formed from the gluon chain of Fig. 2.  They
can be obtained by solving the BFKL equations
$$
\Phi_i (z, k_T^2, Q^2)  =  \Phi_i^{(0)} (z, k_T^2, Q^2)+
$$
\begin{equation}
\overline{\alpha}_S \int_z^1{dz^{\prime}\over 
z^{\prime}} \int {d^2 q\over \pi q^2}
\left[\Phi_i(z^{\prime},(\mbox{\boldmath $q$}+
\mbox{\boldmath $k_T$})^2,Q^2))-\Phi_i(z^{\prime},k_T^2,Q^2)
\Theta(k_T^2-q^2)\right]
\label{eq:a6}
\end{equation}
where $\overline{\alpha}_S \equiv 3 \alpha_S/\pi$.  The
inhomogeneous or driving 
terms $\Phi_{i}^{(0)}$ correspond to the sum of the quark box and
crossed-box 
contributions.  For small $z$ we have
\begin{equation}
\Phi_i^{(0)} (z, k_T^2, Q^2) \; \approx \; \Phi_i^{(0)} (z = 0,
k_T^2, Q^2) \; \equiv \; \Phi_i^{(0)} (k_T^2, Q^2).
\label{eq:aa66}
\end{equation}
We evaluate the $\Phi_i^{(0)}$ by expanding the four momentum in
terms of the 
basic light-like four momenta $p$ and $q^\prime \equiv q + xp$. 
For example, 
the quark momentum ${\kappa}$ in the box (see Fig.~2) has the
Sudakov 
decomposition
$$
\kappa \; = \; \alpha p \: - \: \beta q^\prime \: + \:
\mbox{\boldmath $\kappa$}_T.
$$
We carry out the integration over the box diagrams, subject to
the quark 
mass-shell constraints, and find 
\begin{eqnarray}
\label{eq:a7}
\Phi_T^{(0)} (k_T^2, Q^2) & = & 2 \: \sum_q \: e_q^2 \:
\frac{\alpha _S}{4 \pi^2} \: \frac{Q^2}
{k_T^2} \: \int_0^1 \: d \beta \: \int \:
d^2 \kappa_T \nonumber \\
& & \nonumber \\
& & \left\{ \left[\beta^2 \: + \: (1 - \beta)^2 \right] 
\left(\frac{\kappa_T^2}{D_1^2} \: - \: \frac{\mbox{\boldmath
$\kappa$}_T .
(\mbox{\boldmath $\kappa$}_T - \mbox{\boldmath $k$}_T)}{D_1 D_2}
\right ) \: + \: m_q^2 \left ( \frac{1}{D_1^2} - \frac{1}{D_1
D_2} \right) \right\} \\
& & \nonumber \\
\Phi_L^{(0)} (k_T^2, Q^2) & = & 2 \sum_q \: e_q^2 \;
\frac{\alpha_S}{\pi^2} \frac{Q^4}{k_T^2}
\: \int_0^1 \: d \beta \: \int \:
d^2 \kappa_T \: \beta^2 (1 - \beta)^2 \; \left ( \frac{1}{D_1^2}
\:
- \: \frac{1}{D_1 D_2} \right ). \nonumber 
\end{eqnarray}
where the denominators $D_{i}$ are of the form
\begin{eqnarray}
D_1 & = & \kappa_T^2 \: + \: \beta (1 - \beta) \: Q^2 + m_q^2
\nonumber \\
& & \\
D_2 & = & (\mbox{\boldmath $\kappa$}_T - \mbox{\boldmath
$k$}_T)^2 \:
+ \: \beta (1 - \beta) \: Q^2 + m_q^2 \nonumber.
\label{eq:a8}
\end{eqnarray}
The light $u$, $d$ and $s$ quarks are taken to be massless $(m_q
= 0)$ and
the charm quark to have mass $m_c = 1.4$ GeV. \\

\noindent {\large \bf 2.1  Analytic form of $\Phi$ for
fixed $\mbox{\boldmath$\alpha_S$}$} 
\vspace*{0.3cm} \\
We solve the BFKL equation for $\Phi$ numerically, which allows
the use of running $\alpha_S$ and the inclusion of a charm quark
mass.  However, it is informative to recall the analytic solution
which can be obtained if $\alpha_S$ is fixed and we assume that
the quarks are massless.  The first step is to rewrite the
driving terms (\ref{eq:a7}) for $m_q = 0$ in the form
\begin{eqnarray}
\label{eq:b1}
\Phi_T^{(0)} (k_T^2, Q^2) \; = \; \sum_q \: e_q^2 \:
\frac{\alpha_S}{4 \pi} \: Q^2 \int_0^1 d \lambda \int_0^1 d \beta
\frac{[\beta^2 + (1 - \beta)^2] \: [\lambda^2 + (1 -
\lambda)^2]}{[\lambda (1 - \lambda) k_T^2 + \beta (1 -
\beta) Q^2]}
\end{eqnarray}
\begin{eqnarray}
\label{eq:b2}
\Phi_L^{(0)} (k_T^2, Q^2) \; = \; \sum_q e_q^2 \: \frac{2
\alpha_S}{\pi} \: Q^2 \int_0^1 d \lambda \int_0^1 d \beta
\frac{\lambda (1 - \lambda) \beta (1 - \beta)}{[\lambda (1 -
\lambda) k_T^2 + \beta (1 - \beta) Q^2]}
\end{eqnarray}
where $\lambda$ is the Feynman parameter which appears in the
representation
\begin{equation}
\label{eq:b3}
\frac{1}{D_1 D_2} \; = \; \int_0^1 d \lambda \: \frac{1}{[\lambda
D_1 + (1 - \lambda) D_2]^2}.
\end{equation}
We see that, for fixed $\alpha_S$ and $m_q = 0$, the 
$\Phi_i^{(0)}$ are functions of a single dimensionless variable
$r = Q^2 / k_T^2$.  We may therefore represent the
driving terms $\Phi_i^{(0)} (Q^2 / k_T^2)$ in terms of their
Mellin transforms $\tilde{\Phi}_i^{(0)} (\gamma)$
\begin{equation}
\label{eq:b4}
\Phi_i^{(0)} (r) \; = \; \frac{1}{2 \pi i} \int_{\frac{1}{2} - i
\infty}^{\frac{1}{2} + i \infty} \; d  \gamma \;
\tilde{\Phi}_i^{(0)}
(\gamma) r^{\gamma}
\end{equation}
where $i = L,T$ and $r \equiv Q^2 / k_T^2$.  The Mellin
transform is useful since it diagonalizes the BFKL equation
(\ref{eq:a6}).  The solutions for fixed coupling $\alpha_S$ may
therefore be written
\begin{eqnarray}
\label{eq:b5}
\Phi_i (z, k_T^2, Q^2) \; = \; \frac{1}{2 \pi i}
\int_{\frac{1}{2} - i \infty}^{\frac{1}{2} + i \infty} \; d
\gamma \left(\frac{Q^2}{k_T^2} \right)^{\gamma} {\rm exp}
(\overline{\alpha}_S K(\gamma) \ln \frac{1}{z}) \:
\tilde{\Phi}_i^{(0)} (\gamma)
\end{eqnarray}
where $\overline{\alpha}_S \equiv 3 \alpha_S/\pi$
and $K (\gamma)$ is the Mellin transform of the kernel
of the BFKL equation
\begin{equation}
\label{eq:b6}
K (\gamma) \; = \; 2 \Psi (1) - \Psi (\gamma) - \Psi (1
- \gamma)
\end{equation}
with $\Psi (\gamma) \equiv \Gamma^{\prime} (\gamma) / \Gamma
(\gamma)$.  The functions $\tilde{\Phi}_i^{(0)} (\gamma)$ are
obtained by inserting (\ref{eq:b1}) and (\ref{eq:b2}) into the
inverse relation to (\ref{eq:b4}).  We find 
\begin{eqnarray}
\label{eq:b7}
\tilde{\Phi}_T^{(0)} (\gamma) & = & \sum_q e_q^2 \:
\frac{\alpha_S}{4 \pi} \: \int_0^{\infty} \: dr \: r^{- \gamma}
\:
\int_0^1 d \lambda \int_0^1 d \beta \: \frac{[\beta^2 + (1 -
\beta)^2] [\lambda^2 + (1 - \lambda)^2]}{[\lambda (1 - \lambda) +
\beta (1 - \beta) r]} \nonumber \\
& = & \sum_q e_q^2 \: \frac{\alpha_S}{\sin \pi \gamma} \: B \:
(\gamma +
2, \gamma) B (3 - \gamma, 1 - \gamma)
\end{eqnarray}
\begin{eqnarray}
\label{eq:b8}
\tilde{\Phi}_L^{(0)} (\gamma) & = & \sum_q e_q^2 \: \frac{2
\alpha_S}{\pi} \int_0^{\infty} dr \: r^{- \gamma} \int_0^1 d
\lambda
\int_0^1 d \beta \: \frac{\lambda (1 - \lambda) \beta (1 -
\beta)}{[\lambda (1 - \lambda) + \beta (1 - \beta) r]} \nonumber
\\
& = & \sum_q e_q^2 \: \frac{2 \alpha_S}{\sin \pi \gamma} \: B
(-
\gamma + 2, - \gamma + 2) B (\gamma + 1, \: \gamma + 1)
\end{eqnarray}
where $B(x,y) \equiv \Gamma(x) \Gamma(y) / \Gamma(x+y)$.
The derivation of the analytic formula relies on $\alpha_S$ being
fixed.  This approach has been used by Bartels et al.
\cite{BARTELS} to estimate the DIS + forward jet cross section
taking the coupling $\alpha_S (k_T^2)$ in
formulae (\ref{eq:b5}). The prediction has the general
shape of the H1 data as a function of $x$, but the calculated
cross section exceeds the data by some 20\% \cite{H1data}.  

In the $z \rightarrow 0$ limit the formulae reduce to the
conventional $z^{- \lambda}$ BFKL
behaviour
\begin{eqnarray}
\label{eq:b9}
\Phi_i (z, k_T^2, Q^2) \sim z^{- \overline{\alpha}_S K
(\frac{1}{2})} \left(\frac{Q^2}{k_T^2} \right)^{\frac{1}{2}}
\; \frac{\tilde{\Phi}_i^{(0)} (\gamma =
\frac{1}{2})}{(\overline{\alpha}_S K^{\prime \prime}(\frac{1}{2})
\ln
1/z)^{\frac{1}{2}}}
\end{eqnarray}
where for simplicity we have omitted the Gaussian diffusion
factor in $\ln (k_T^2 / Q^2)$.  If we evaluate the various
functions at $\gamma = \frac{1}{2}$ we obtain
\begin{eqnarray}
\label{eq:b10}
\Phi_T (z, k_T^2, Q^2) & = & \frac{9 \pi^2}{512} \;
\frac{2
\sum e_q^2 \: \alpha_S^{\frac{1}{2}}}{\sqrt{21 \zeta (3) / 2}}
\left(\frac{Q^2}{k_T^2} \right)^{\frac{1}{2}} \: \frac{z^{-
\lambda}}{\sqrt{\ln (1/z)}} \left[1 + O \left(\frac{1}{\ln (1/z)}
\right) \right] \nonumber \\ 
\Phi_L (z, k_T^2, Q^2) & = &
\frac{2}{9} \: \Phi_T (z, k_T^2, Q^2)
\end{eqnarray}
where $\lambda = \overline{\alpha}_S \: K (\frac{1}{2}) =
\overline{\alpha}_S \: 4 \ln 2$. \\

\noindent {\large \bf 2.2 Normalisation of $\Phi$ and the
description of DIS + jet data} 
\vspace*{0.3cm} \\
Our calculation of the DIS + forward jet process differs from
that of ref.\ \cite{BARTELS} in that we numerically solve the
BFKL equations.  Therefore, we are able to explicitly include the $m_c
\neq 0$ charm contribution.  We also allow the coupling $\alpha_S$ to
run. To be precise we solve the BFKL equation (\ref{eq:a6}) rewritten
in terms of the modified function $\overline{\alpha}_S (k_{jT}^2)
\Phi_i (z,k_T^2,Q^2)$ following the prescription that was used in
ref.~\cite{KMS1}.  This choice
of scale for $\alpha_S$ is consistent with the double logarithm limit and 
with the NLO $\ln (1/x)$ analysis of ref.~\cite{CIAF}.

We determine the functions $\Phi_i$ for $z < z_0$ by solving the
BFKL equation as described in \cite{KMS1} starting from the
boundary condition
\begin{equation}
\label{eq:b11}
\Phi_i (z_0, k_T^2, Q^2) \; = \; \Phi_i^{(0)} (z_0,
k_T^2, Q^2) \: \approx \: \Phi_i^{(0)} (k_T^2, Q^2)
\end{equation}
where $\Phi_i^{(0)} (k_T^2, Q^2)$ are the contributions of
the quark box (and crossed box) given in (\ref{eq:a7}).  We take
$u,d,s$ to
be massless and the charm quark to have mass $m_c = 1.4 {\rm
GeV}$ in the summation over the quarks.  We then use
(\ref{eq:a2}) integrated over $x$ and $Q^2$ 
and (\ref{eq:a3}) to calculate the DIS + forward
jet rate corresponding to the cuts used in the H1 measurement. 
That is the forward jet is constrained to the region
$$7^{\circ} < \theta_j < 20^{\circ}, \quad E_j > 28.7 {\rm GeV},
\quad
k_{jT} > 3.5 {\rm GeV},$$
whereas  the outgoing electron must lie in the domain
$$160^{\circ} < \theta_e^{\prime} < 173^{\circ}, \quad
E_e^{\prime} > 11 {\rm GeV}, \quad y > 0.1$$
in the HERA frame.  Finally H1 require $\frac{1}{2} Q^2 <
k_{jT}^2 < 2Q^2$.  The BFKL calculation is compared with the
data \cite{H1data} in bins of size $\Delta x \: = \: 5 \times
10^{-4}$ in
Fig.~3. The parameter $z_0$ is adjusted to give a satisfactory
normalization of the calculation.  We find that the H1 data
require $z_0 = 0.15$.  The predicted shape of the distribution is
in good agreement with the data. \\

\noindent {\large \bf 3.  Single particle $\mbox{\boldmath$p_T$}$
spectra} 
\vspace*{0.3cm} 

We first use Fig.~1(c) to obtain the differential cross section
for the
production of a hadron of transverse momentum $p_T$ and
longitudinal
momentum fraction $x_h$. Then we calculate the charged particle
spectra
relevant to the recent observations at HERA \cite{H1ptdata}.\\

\noindent {\large \bf 3.1  The cross section for charged particle
production} 
\vspace*{0.3cm} 

The cross-section for single particle production is
obtained by convoluting the inclusive cross-section for the
production of a single parton with the parton fragmentation
function.  The differential cross section for the inclusive
production of a single parton of longitudinal momentum fraction
$x_j$ and transverse momentum $k_{jT}$ has the generic form
of (\ref{eq:a2}).  We have
\begin{eqnarray}
\label{eq:b12}
\frac{\partial \sigma_j}{\partial x_j \: \partial k_{jT}^2 \:
\partial x \: \partial Q^2} \; = \; \frac{4 \pi \alpha^2}{x Q^4}
\left[ (1 - y) \frac{\partial F_2}{\partial x_j \: \partial
k_{jT}^2} \: + \: \frac{1}{2} \: y^2 \: \frac{\partial
F_T}{\partial x_j \: \partial k_{jT}^2} \right].
\end{eqnarray}

Now for small $x$, and in the central region away from the
current jet and the proton remnants, we expect gluonic partons to
dominate where the gluons
are radiated within the BFKL ladder.  The differential
structure functions occurring in (\ref{eq:b12}) are then given by
\begin{eqnarray}
\label{eq:b13}
x_j \: \frac{\partial F_i}{\partial x_j \: \partial k_j^2}
\; = \; \int \frac{d^2 k_p}{\pi k_p^4} \int \frac{d^2
k_{\gamma}}{k_{\gamma}^2} \left[ \frac{\overline{\alpha}_S
(k_j^2) k_p^2 k_{\gamma}^2}{k_j^2} \right] f (x_j, k_p^2) \:
\Phi_i \left(\frac{x}{x_j}, k_{\gamma}^2, Q^2 \right) \delta^2
(k_j - k_p - k_{\gamma})
\end{eqnarray}
with $i = T,L$ and where for simplicity we have omitted the
subscript $T$ from the gluon transverse momenta, $k_{jT}, \:
k_{pT}$ and $k_{\gamma T}$, see Fig.~4.  The functions
$\Phi_i$ are those of Sec.~2 which control the DIS + forward
jet rate, while $f$ is the unintegrated gluon distribution which
satisfies the BFKL equation
\begin{eqnarray}
\label{eq:b14}
-z \: \frac{\partial f (z, k^2)}{\partial z} \; = \;
\overline{\alpha}_S \int \frac{d^2 q}{\pi q^2}
\left[\frac{k^2}{(\mbox{\boldmath $q$} + \mbox{\boldmath
$k$})^2} \: f (z, (\mbox{\boldmath $q$} + \mbox{\boldmath
$k$})^2) \: - \: f (z, k^2) \Theta (k^2 - q^2)\right].
\end{eqnarray}
The expression in square brackets in (\ref{eq:b13}) arises from
the (square of the) BFKL vertex for real gluon emission, see
Fig.~4.

In practice we evolve (\ref{eq:b14}) down in $z$ from the
boundary condition
\begin{equation}
\label{eq:b15}
f (\overline{z}_0, k^2) \; = \; f^{AP} (\overline{z}_0, k^2) \; =
\; \frac{\partial \left[ \overline{z}_0 \: g^{AP}
(\overline{z}_0, k^2) \right]}{\partial \ln (k^2/k_0^2)}
\end{equation}
where here $z = \overline{z}_0$ with $\overline{z}_0 = 10^{-2}$, 
and where $g^{AP}$ is
the conventional gluon distribution obtained from a global set of
partons.  As before we allow the coupling to run, that is we take
$\alpha_S (k^2)$ in (\ref{eq:b14}).  Moreover, we impose an
infrared cut-off $k_0^2 = 1 {\rm GeV}^2$.  That is we
require the arguments of $f$ to satisfy $k^2 > k_0^2$
and $(\mbox{\boldmath $k$} + \mbox{\boldmath $q$})^2 >
k_0^2$. Similarly, the integrations in (\ref{eq:b13}) are
restricted to the regions $k_p^2, \: k_{\gamma}^2 >
k_0^2$.  We may include the contribution $\Delta F_i$ from
the region $k_p^2 < k_0^2$ by assuming the strong
ordering approximation, $k_p^2 \ll k_{\gamma}^2 \sim
k_j^2$, at the gluon vertex.  This contribution to
(\ref{eq:b13}) then becomes
\begin{eqnarray}
\label{eq:b16}
x_j \: \frac{\partial (\Delta F_i)}{\partial x_j \partial
k_j^2}
& = & \overline{\alpha}_S (k_j^2) \int^{k_0^2}
\frac{d k_p^2}{k_p^2} \: f (x_j, k_p^2) \; \Phi_i
\left( \frac{x}{x_j}, \: k_j^2, \: Q^2 \right) \nonumber \\ 
& = &
\overline{\alpha}_S (k_j^2) \: \frac{x_j g (x_j, k_0^2)}{k_j^2}
\; \Phi_i \left( \frac{x}{x_j}, k_j^2, Q^2 \right).
\end{eqnarray}
Most of the time, however, for the calculation relevant to the
HERA data, the variable $x_j$ is not small enough for the BFKL
equation to be applicable for the function $f$.  In these cases,
that is when $x_j > \overline{z}_0$, we therefore again assume
strong
ordering $k_p^2 \ll k_{\gamma}^2 \sim k_j^2$. In addition we
include
the contributions from quark and antiquark jets. We then obtain
\begin{equation}
\label{eq:b17} 
x_j \; \frac{\partial F_2}{\partial x_j \partial k_j^2} \; = \;
\overline{\alpha}_S (k_j^2) \; \frac{x_j \left[ g + \frac{4}{9}
\sum_q \left( q + \bar{q} \right) \right] }{k_j^2}
\; \Phi_i \left(\frac{x}{x_j}, \: k_j^2, \: Q^2 \right).
\end{equation}
where the parton distributions are to be evaluated at $(x_j,
k_j^2)$
The differential cross section for single particle $(h)$
production is obtained by convoluting the jet cross section with
the fragmentation functions $D$ for the parton $\rightarrow h$
transition
\begin{eqnarray}
\label{eq:b18}
\frac{\partial \sigma_h}{\partial x_h \partial p_T^2 \partial x
\partial Q^2} & = & \int_{x_h}^1 \; dz \; \int \; dx_j \; \int
\; dk_j^2 \: \delta (x_h - z x_j) \delta (p_T - zk_j) \left\{
\frac{\partial \sigma_g}{\partial x_j \partial k_j^2
\partial x \partial Q^2} \: D_g^h (z, \mu^2) \right. \nonumber \\
& + & \left. \frac{4}{9} \sum_q \left[ \frac{\partial
\sigma_q}{\partial x_j \partial
k_j^2 \partial x \partial Q^2} D_q^h (z, \mu^2)  \; + \;
\frac{\partial \sigma_{\overline{q}}}{\partial x_j \partial k_j^2
\partial x \partial Q^2} D_{\overline{q}}^h (z, \mu^2) \right]
\right\}
\end{eqnarray}
where $\sigma_g$ and $\sigma_q$ are the contributions to the
cross section
$\sigma_j$ for gluon and quark and antiquark jets respectively. 
The fragmentation scale $\mu^2$ is of the order of $k_j^2$.  The
cross
section for charged particle production is obtained by summing
over all possible charged hadrons $h$.\\

\noindent {\large \bf 3.2  Predictions for the single particle 
$\mbox{\boldmath$p_T$}$ spectra} 
\vspace*{0.3cm} 

The data for the single (charged) particle $p_T$ spectra are
presented
in the form $(dn/dp_T)/N$ where $n$ is the multiplicity 
and $N$ the total number of charged particles in a given $x, Q^2$
bin
\cite{H1data}. To calculate this $p_T$ spectrum we evaluate
\begin{equation}
\label{eq:b19}
\frac{1}{N} \: \frac{dn}{dp_T} \: = \: \left( \sum_h
\frac{\partial
\sigma_h}{\partial p_T \: \partial x \: \partial Q^2} \right)
\left/
\frac{\partial \sigma_{tot}}{\partial x \: \partial Q^2} \right. .
\end{equation} 
where $\partial \sigma_h / \partial p_T \partial x \partial Q^2$
is
obtained from (\ref{eq:b18}) by integrating over $x_h$. We take
the
central values of $x, Q^2$ in the bin. 
The integration limits are fixed by the limits on
the pseudorapidity interval under consideration. To be precise we
use
\begin{equation}
\label{eq:b20}
x_h \: = \: \sqrt{\frac{x}{Q^2}} \: p_T \: e^{-\eta}
\end{equation}
where $\eta$ is the pseudorapidity of the charged particle,
$\eta = - \ln \tan (\theta / 2)$ with $\theta$ the angle with
respect
to the virtual photon direction. Finally we calculate the total
differential
cross section $\partial \sigma_{tot} / \partial x \partial Q^2$ 
in (\ref{eq:b19})
from the structure functions $F_2$ and $F_L$ given by the MRS(R2)
\cite{mrs} set
of parton distributions.

Our aim is to make an absolute BFKL-based prediction to compare
with the
$p_T$ spectra observed by the H1 collaboration. There is,
however, an
inherent uncertainty in the normalisation due to the imposition
of
an infrared cut-off on the BFKL transverse momentum integrations
(or due to other possible treatments of the non-perturbative
region).
To overcome this problem we follow the
procedure described in Sec.~2.2 and fix the parameters
occurring
in the calculation of the BFKL functions $\Phi_i$ by requiring
the
prediction for the DIS + jet cross section to give the correct
normalization of the H1 forward jet measurements. The next step
is to use  
the functions $\Phi_i$ obtained in this way in the computation of
the
differential structure functions from (\ref{eq:b13}),
(\ref{eq:b16})
and (\ref{eq:b17}). In this way we are able to calculate a
normalized 
$p_T$ spectrum from (\ref{eq:b19}).

The BFKL prediction for the single particle spectra may be
compared with the result which would be obtained if the BFKL
gluon radiation
is neglected. That is in (\ref{eq:b13}), (\ref{eq:b16}) and
(\ref{eq:b17})
we replace the functions $\Phi_i$ which describe
the solution of the BFKL equation with the boundary condition
given
by the quark box with the quark box $\Phi_i^{(0)}$ only. In
addition 
we now also assume strong ordering for $x_j < \overline{z}_0$ and
carry out the $k_p^2$ integration in (\ref{eq:b13}).  This
amounts to assuming that in a fixed-order treatment the dominant
subprocess is $\gamma g \rightarrow q \overline{q} g$.  In our
calculation the $\kappa$ integration is infrared finite since we
allow for the virtuality of the incoming and exchanged gluons. 

So we are now in the position to give a BFKL prediction for the
single
particle spectra which can be compared with the H1 data. 
In their measurement the H1 collaboration collected
data in nine different kinematic bins in two pseudorapidity
intervals.
We will focus on the three smallest $x$ bins where BFKL effects
should
become visible. Also we will only show results for the lower
pseudorapidity
interval, $0.5 < \eta < 1.5$, where we expect no contamination
due to
the fragmentation of the current jet which has not been included
in the
calculation. In the computation of the $p_T$ spectra we use
$$E_e = 27.5 {\rm GeV}, \quad E_p = 820 {\rm GeV}$$
and impose the cuts
which where used in the H1 measurement, i.e.\ we require the
outgoing
electron to lie in the region
$$157^{\circ} < \theta_e^{\prime} < 173^{\circ}, \quad
E_e^{\prime} > 12 {\rm GeV}, \quad y > 0.05$$
in the HERA frame. Also we subtract 10\% off the total cross
section
$\sigma_{tot}$ to account for diffractive events with large
rapidity
gaps which have been excluded from the measurement. Finally, in
the sum
over the charged hadrons $h$ in (\ref{eq:b19}) we include
$\pi^{\pm}$ 
and $K^{\pm}$, and we use the next-to-leading order fragmentation functions by
Binnewies 
et al.\ \cite{BKK}. In Fig.~5 we show predictions for the charged
particle $p_T$ spectrum in kinematic bin 1 of the H1 analysis
with central values $x = 1.6 \times 10^{-4}$ and $Q^2 = 7$
GeV$^2$. We compare
the results when BFKL small $x$ resummation is included in the
calculation 
with the case when
gluon radiation is neglected. In both cases we demonstrate the
effect of
changing the fragmentation scale from $\mu^2 = k_j^2$ to 
$\mu^2 = (2k_j)^2$.  We see that the BFKL prediction gives a good
description of both the shape and the normalization 
\footnote{Even though we have normalised $\Phi$ to the DIS + jet
data, there still remains some residual uncertainty in the overall
normalisation associated with the choice of infrared cut-off used
in the $k_\gamma$ integration in (\ref{eq:b13}). Our results are
shown for the natural choice $k_{\gamma 0}^2 = 1$ GeV$^2$.}
of the H1
data.  On the other hand, when the BFKL effects are neglected the
predictions lie considerably below the data.  Also we see, as
expected, that the spectrum decreases more rapidly with
$p_T$ than when the BFKL resummation is included. For example for
$p_T = 1.5$ GeV the two predictions differ by a factor 3.6,
whereas for $p_T = 6$ GeV
this factor is almost 10. This is a reflection of the diffusion
in $\ln k_T^2$ along the BFKL ladder.  

The same general behaviour is seen in Figs.~6 and 7 where we show
the comparison for kinematic
bins 2 and 3, with central values $x = 2.9 \times 10^{-4}$, $Q^2
= 9$ GeV$^2$
and $x = 3.7 \times 10^{-4}$, $Q^2 = 13$ GeV$^2$, respectively.
We find that in
all three small $x$ bins of the H1 analysis the data strongly
support
the inclusion of BFKL resummation in the calculation of the $p_T$
spectra. Reasonable variations of the fragmentation scale do not
allow
for a description of the data when BFKL effects are neglected.\\

\noindent {\large \bf 4.  Conclusion}
\vspace*{0.3cm} 

We studied the DIS + forward jet process including massive charm
in the
quark box and solving the BFKL equation numerically for running
coupling. We found that BFKL
dynamics describe the shape of the $x$ distribution of the HERA
data
well. Next we used these data to fix the normalization of the
solution
of the BFKL equation with the boundary condition given by the
quark
box. This enabled us to give an absolute prediction for charged
particle transverse momentum spectra at small $x$. We calculated
the spectrum
for large values of $p_T$ first including BFKL small $x$
resummation
in the calculation and second neglecting gluon radiation. It
turned
out that the BFKL prediction agrees well with the H1 data both in
shape
and normalization, whereas the approximate fixed order result 
underestimates the data and decreases too rapidly with $p_T$.
We therefore conclude that we found evidence for the existence of
$\ln(1/x)$ effects and for the diffusion in $\ln k_T^2$ which
accompanies 
BFKL evolution. Despite these encouraging results it would,
however, still be 
useful to compare the BFKL prediction for the $p_T$ spectrum
with the result of the 
complete fixed order calculation. Experimental data for higher
values of $p_T$
would allow an even clearer distinction between the different
predictions.
BFKL effects would also become more apparent in the 
pseudorapidity interval $-0.5 < \eta < 0.5$ which corresponds to 
higher values of $x_j$ and therefore to a longer BFKL evolution
starting from
the quark box. Of course higher $x_j$ also means less BFKL
evolution from the
proton end. This is, however, not a disadvantage, since already
for the 
pseudorapidity interval which we 
considered the main contribution to the spectrum comes from the
region $x_j >
\overline{z}_0$. We conclude that although more experimental data
especially for
higher values of $p_T$ would be useful, the existing spectra show
the presence of BFKL effects at small $x$ at HERA.\\ 

\noindent {\large \bf Acknowledgements}
\vspace*{0.3cm}

We thank Michael Kuhlen and Erwin Mirkes for their help and
encouragement.  S.C.L. thanks the UK Engineering and Physical
Sciences Research Council for financial support.  This work has
been supported in part by Polish State Committee for Scientific
Research Grant No. 2 P03B 089 13, and by the EU
under Contracts Nos. CHRX-CT92-0004 and CHRX-CT93-357.



\newpage
\noindent {\large \bf Figure Captions}
\begin{itemize}

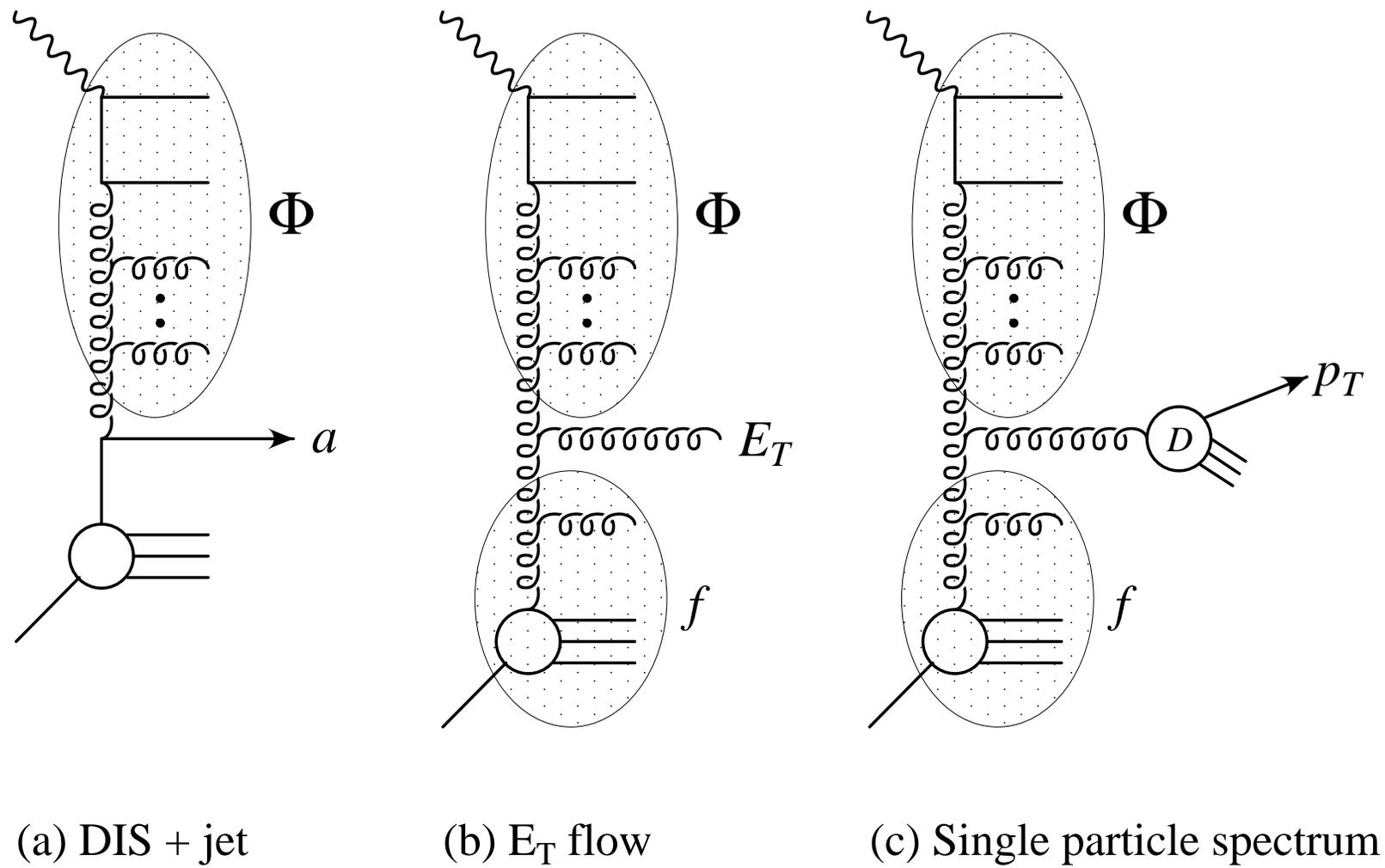
\item[Fig.\ 1] Diagrammatic representation of (a) the deep
inelastic + forward
jet, (b) the $E_T$ flow, and (c) the single particle spectrum
measurement.

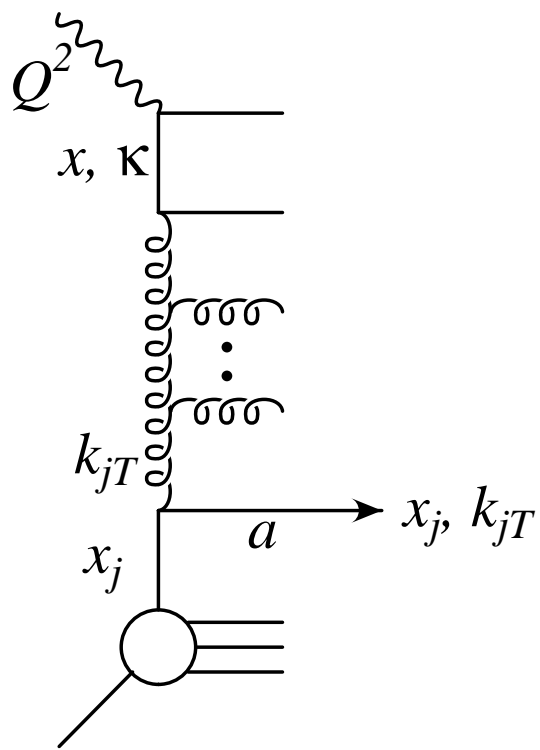
\item[Fig.\ 2] Diagrammatic representation of a deep inelastic +
forward
jet event.

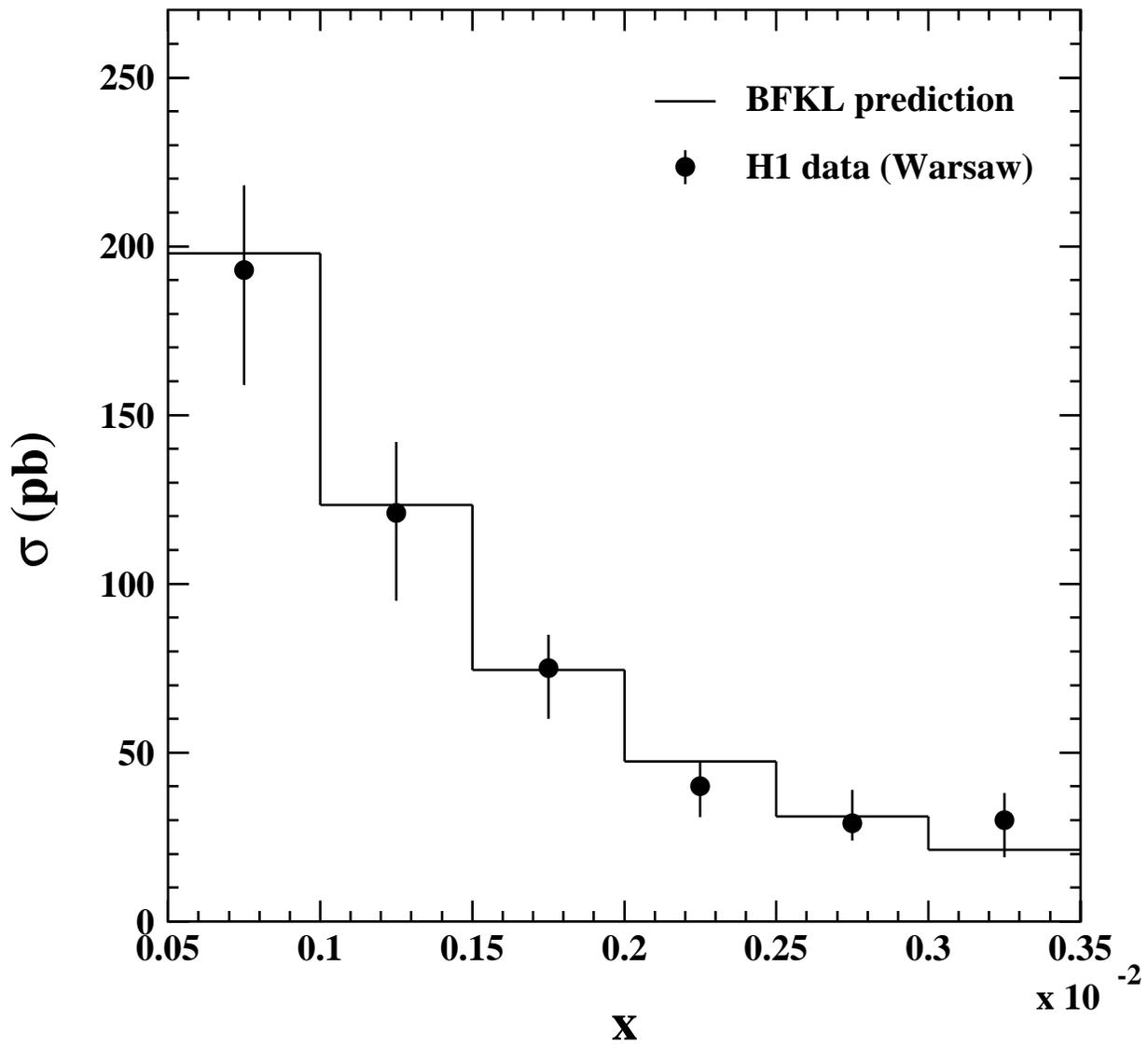
\item[Fig.\ 3] The deep inelastic + forward jet cross section in
pb
integrated over bins of size $5 \times 10^{-4}$ in $x$ compared
to
the H1 data presented at the Warsaw conference \cite{H1data}. 
As in the H1 measurement the forward jet was required to fulfil
$7^{\circ} < \theta_j < 20^{\circ}$, $E_j > 28.7$ GeV, and
$k_{jT} > 3.5$ GeV. The electron acceptance region is limited by
$160^{\circ} < \theta_e^{\prime} < 173^{\circ}$,
$E_e^{\prime} > 11$ GeV, and $y > 0.1$ in the HERA frame.

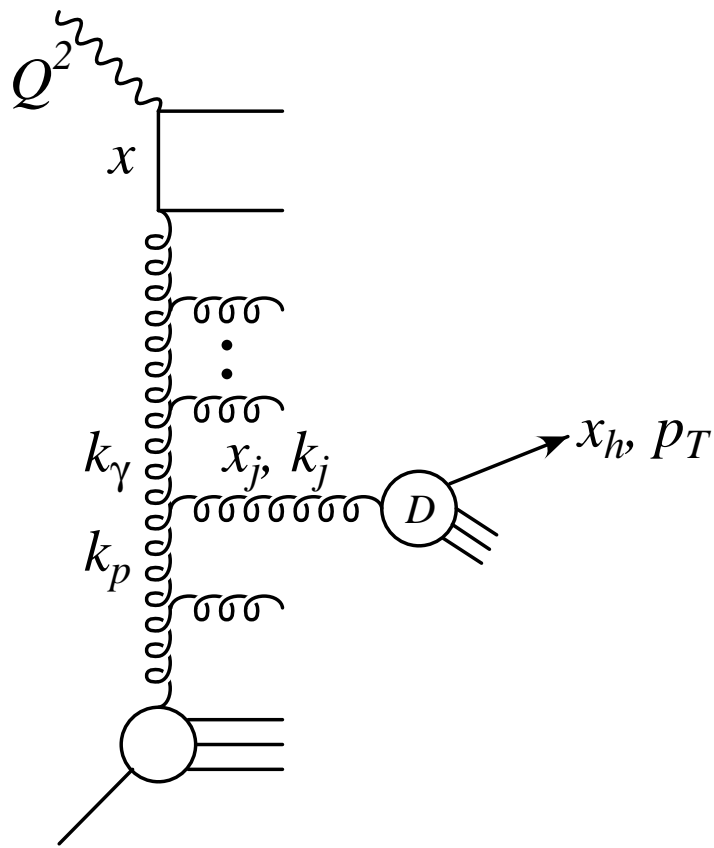
\item[Fig.\ 4] Diagrammatic representation of the cross section
for
emission of a high transverse momentum $p_T$ particle.

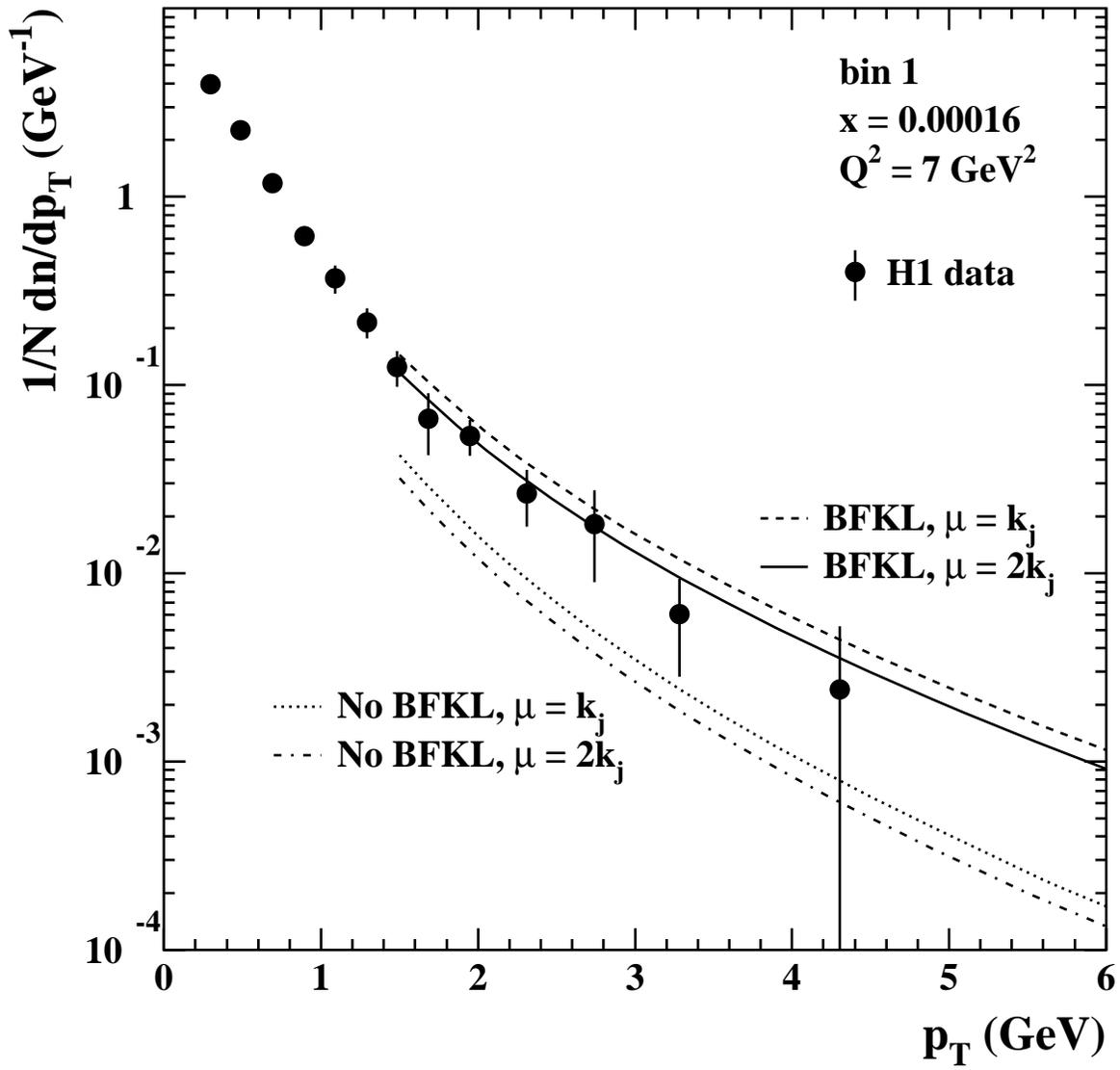
\item[Fig.\ 5] The transverse momentum spectrum of charged
particles
($\pi^+, \pi^-, K^+, K^-$) in the pseudorapidity interval 
$0.5 < \eta < 1.5$ in the virtual photon-proton centre-of-mass
frame.
The results are shown for kinematic bin 1 with the central values
$x = 1.6 \times 10^{-4}$ and $Q^2 = 7$ GeV$^2$. The continuous
and the
dashed curve show the spectra obtained with $\Phi_i$ and $f$ 
calculated from the BFKL equation. They only differ in the choice
of
fragmentation scale: for the continuous curve the fragmentation
functions were evaluated at scale $\mu^2 = (2k_j)^2$ and for the
dashed curve
at scale $\mu^2 = k_j^2$. When BFKL radiation is neglected in the
calculation
of the $p_T$ spectra, i.e.\ when the quark box approximation 
$\Phi_i = \Phi_i^{(0)}$ is used and strong ordering at the gluon
vertex
is assumed, then the
dash-dotted and dotted curves are obtained. The fragmentation
functions
were evaluated at scales $2k_j$ and $k_j$ respectively. The data
points
shown are from the H1 measurement of the charged particle spectra
\cite{H1ptdata}.

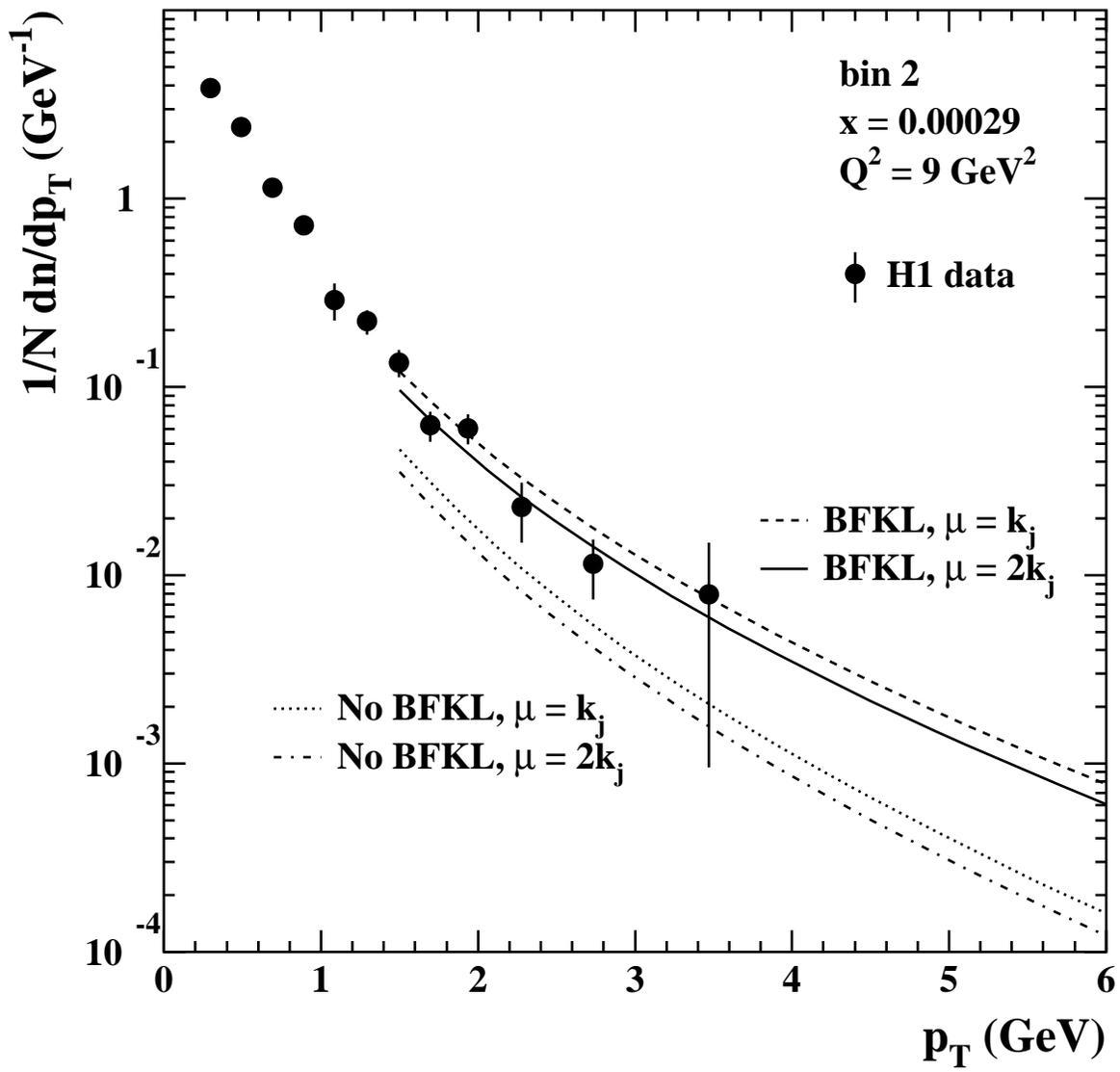
\item[Fig.\ 6] As Fig.\ 5, but for kinematic bin 2, 
$x = 2.9 \times 10^{-4}$ and $Q^2 = 9$ GeV$^2$.

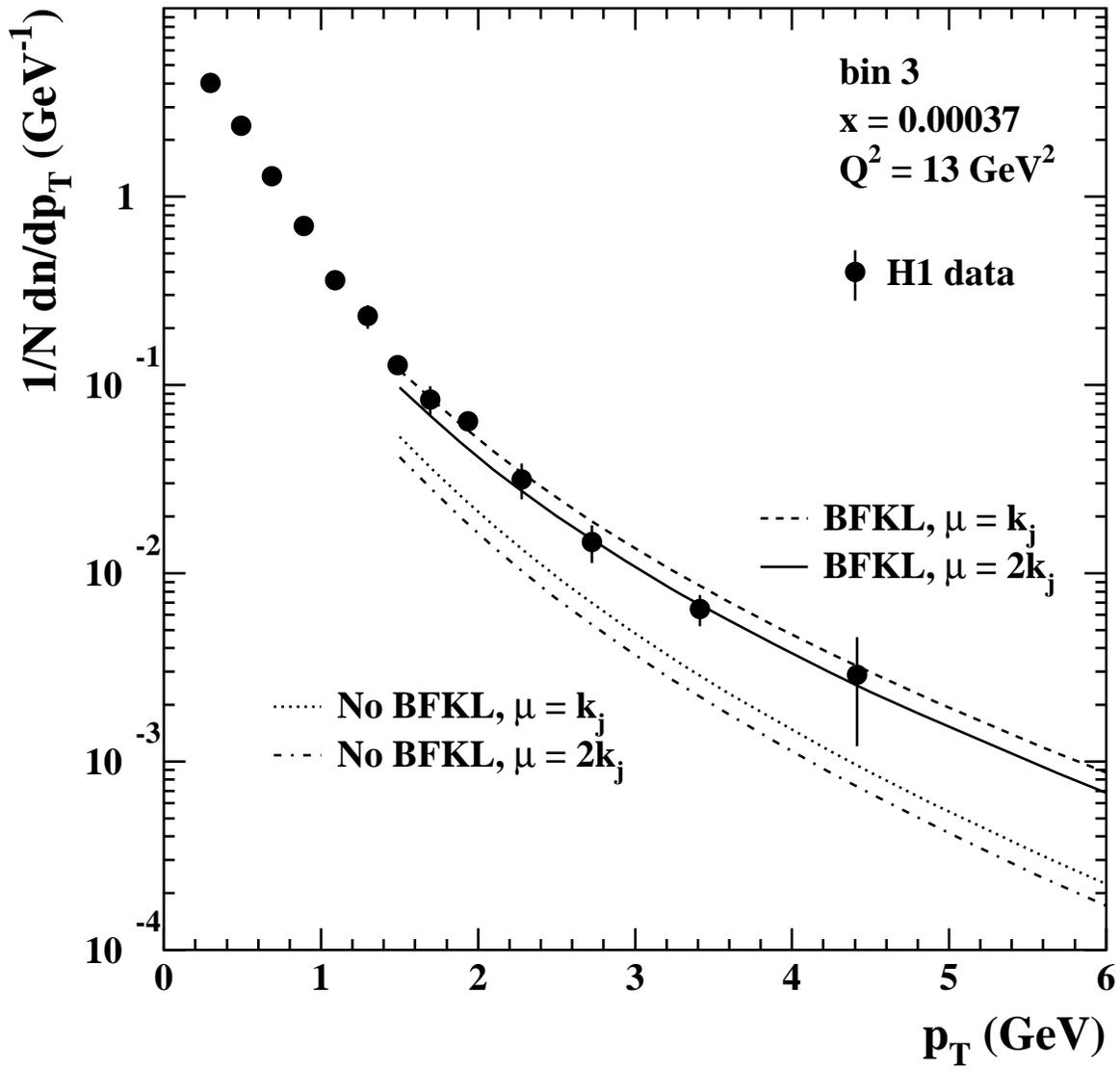
\item[Fig.\ 7] As Fig.\ 5, but for kinematic bin 3, 
$x = 3.7 \times 10^{-4}$ and $Q^2 = 13$ GeV$^2$.

\end{itemize} 


\newpage
\begin{thebibliography}{xx}
\bibitem{BFKL} E.\ A.\ Kuraev, L.\ N.\ Lipatov and V.\ Fadin,
Zh.\ Eksp.\ Teor.\ Fiz.\ {\bf 72}, 373 (1977) (Sov.\ Phys.\ JETP
{\bf 45}, 199 (1977)); \\
Ya.\ Ya.\ Balitzkij and L.\ N.\ Lipatov, Yad.\ Fiz.\ {\bf 28},
1597 (1978) (Sov.\ J.\ Nucl.\ Phys.\ {\bf 28}, 822 (1978)); \\
L.\ N.\ Lipatov, in \lq\lq Perturbative QCD", edited by A.\ H.\
Mueller, (World Scientific, Singapore, 1989), p. 441; \\
J.\ B.\ Bronzan and R.\ L.\ Sugar, Phys.\ Rev.\ {\bf D17}, 585
(1978); \\
T.\ Jaroszewicz, Acta.\ Phys.\ Polon.\ {\bf B11}, 965 (1980).
%
\bibitem{STASTO} J.\ Kwiecinski, A.\ D.\ Martin and A.\ M.\
Stasto,
Durham preprint DTP/97/18, Phys.\ Rev. (in press).
%
\bibitem{DGLAP} Yu.\ Dokshitzer, Soviet Phys.\ JETP {\bf 46}
(1977) 641;\\
V.\ N.\ Gribov and L.\ N.\ Lipatov, Soviet J.\ Nucl.\ Phys.\ {\bf
15} (1972) 
438, 675;\\
G.\ Altarelli and G.\ Parisi, Nucl.\ Phys.\ {\bf B126} (1977)
298.
%
\bibitem{M} A.\ H.\ Mueller, Nucl.\ Phys.\ B (Proc.\ Suppl.) {\bf
18C} (1990) 125; J.\ Phys.\ {\bf G17} (1991) 1443; \\
W.\ K.\ Tang, Phys.\ Lett.\ {\bf B278} (1992) 363; \\
J.\ Bartels, A.\ De Roeck and M.\ Loewe, Z.\ Phys.\ {\bf C54}
(1992) 635; \\
A.\ De Roeck, Nucl.\ Phys.\ B (Proc.\ Suppl.) {\bf 29A} (1992)
61; \\
J.\ Kwiecinski, A.D.\ Martin and P.J.\ Sutton, Phys.\ Lett.\ {\bf
B287} (1992) 254;  Nucl.\ Phys.\ B (Proc.\ Suppl.) {\bf 29A}
(1992) 67.
%
\bibitem{KMS1} J.\ Kwiecinski, A.D.\ Martin and P.J.\ Sutton, 
Phys.\ Rev.\ {\bf D46} (1992) 921.
%
\bibitem{gkms} K.\ Golec-Biernat, J.\ Kwiecinski, A.\ D.\ Martin
and P.\ J.\ Sutton, Phys.\ Rev.\ {\bf D50} (1994) 217; Phys.\
Lett.\ {\bf B335} (1994) 220.
%
\bibitem{ETflow} H1 collaboration: S.\ Aid {\it et al.}, Phys.\
Lett.\ {\bf B356} (1995) 118. 
%
\bibitem{MK} M.\ Kuhlen, Phys.\ Lett.\ {\bf B382} (1996) 441;
Contribution to the Workshop on \lq\lq Future Physics at HERA",
Hamburg 1996, hep-ex/9610004.
%
\bibitem{BARTELS} J.\ Bartels, V.\ Del Duca, A.\ De Roeck, D.\
Graudenz,
and M.\ W\"{u}sthoff, Phys.\ Lett.\ {\bf B384} (1996) 300. 
%
\bibitem{H1data} H1 Collaboration, C.\ Adloff {\it et al.},
contributed paper pa03-049, ICHEP '96, Warsaw, Poland, July
1996.
%
\bibitem{CIAF} M.\ Ciafaloni and G.\ Camici, contribution to the
Ringberg Workshop:
New Trends in HERA Physics, May 1997. 
%
\bibitem{H1ptdata} H1 Collaboration, C.\ Adloff {\it et al.},
Nucl.\ Phys.\ {\bf B485} (1997) 3.
%
\bibitem{mrs} A.\ D.\ Martin, R.\ G.\ Roberts and W.\ J.\
Stirling,
Phys.\ Lett.\ {\bf B387} (1996) 419.
%
\bibitem{BKK} J.\ Binnewies, B.\ A.\ Kniehl and G.\ Kramer,
Phys.\ Rev.\ {\bf D52} (1995) 4947.
\end{thebibliography}
\end{document}